\documentclass[aps,prb,twocolumn,superscriptaddress,floatfix,longbibliography]{revtex4-2}

\usepackage[english]{babel}
\usepackage[utf8]{inputenc}
\usepackage{mathtools}

\usepackage{physics}
\usepackage{float}
\usepackage{braket}
\usepackage{bbm}
\usepackage{bm}
\usepackage{amsbsy}
\usepackage{amsthm}
\usepackage{amssymb}
\usepackage{amsfonts}
\usepackage{amsmath}
\usepackage{soul}
\setstcolor{red}
\usepackage{dsfont} 
\usepackage{graphicx} 
\usepackage{hhline}
\usepackage{epsfig}
\usepackage{epstopdf}
\usepackage{dsfont}
\usepackage{xcolor}
\usepackage{color}
\usepackage{verbatim}
\usepackage{nomencl}
\usepackage[colorlinks]{hyperref}
\usepackage{float}
\usepackage[normalem]{ulem}
\usepackage{multirow}
\usepackage{makecell}
\usepackage{url}
\DeclareUnicodeCharacter{0301}{\'{e}}

\begin{document}
\title{Prethermal Floquet time crystals in chiral multiferroic chains and applications as quantum sensors of AC fields}
\author{Rohit Kumar Shukla}
\email[]{rohitkrshukla.rs.phy17@itbhu.ac.in}
\affiliation{Optics and Quantum Information Group, The Institute of Mathematical Sciences, CIT Campus, Taramani, Chennai 600113, India}
\affiliation{Homi Bhabha National Institute, Training School Complex, Anushakti Nagar, Mumbai 400085, India}
\affiliation{Department of Chemistry; Institute of Nanotechnology and Advanced Materials; Center for
Quantum Entanglement Science and Technology, Bar-Ilan University, Ramat-Gan, 5290002 Israel}
\author{Levan Chotorlishvili}
\affiliation{Department of Physics and Medical Engineering, Rzeszow University of Technology, 35-959 Rzeszow Poland}
\author{Sunil K. Mishra}
\affiliation{Department of Physics, Indian Institute of Technology (Banaras Hindu University) Varanasi - 221005, India}
\author{Fernando Iemini}
\affiliation{Instituto de Fı́sica, Universidade Federal Fluminense, 24210-346 Niterói, Brazil}
\date{\today}
\begin{abstract}
We study the emergence of prethermal Floquet Time Crystal (pFTC) in disordered chiral multiferroic chains. The model is an extension of the usual periodically driven nearest-neighbor disordered Heisenberg chain, with additional next-nearest-neighbor Heisenberg couplings and DMI interactions due to external magnetic and electric couplings. We derive the phase diagram of the model, characterizing the magnetization, entanglement, and coherence dynamics of the system along the extended interactions. In addition, we explore the application of the pFTC as quantum sensors of AC fields. The sensor performance to estimate small AC fields is quantified through the quantum Fisher information (QFI) measure. The sensor offers several advantages as compared to those composed of non-interacting spins due to its intrinsic robustness, long coherent interrogation time, and many-body correlations. Specifically, the sensor can overcome the standard quantum limit ($\rm{SQL} \sim  N t^2$) during the prethermal regime, reaching an optimum performance at the pFTC lifetime $t^*$, where the $\rm{QFI}/Nt^{*^2} \sim  N^\alpha$ with $\alpha > 0$, scaling superlinarly with the number of spins. Different from \text{full} FTCs, the prethermal lifetime does not diverge in the thermodynamic limit, nevertheless it can be    increasingly long with tuning system parameters.
 \end{abstract}


\maketitle

\section{Introduction}
The study of quantum phenomena and their potential applications has led to the emergence of novel concepts in condensed matter physics. Among them, time crystals have attracted considerable attention due to their unique nonequilibrium properties. A time crystal, originally proposed by Wilczek in isolated systems \cite{wilczek_quantum_2012}, is a distinct phase of matter characterized by the spontaneous breaking of time translation symmetry in many-body systems. Soon after its proposal, a no-go theorem was established for energy-conserving systems \cite{watanabe2015absence,bruno2013impossibility}, leading to the idea that the appropriate context for such phases lies in systems operating out of equilibrium.

The most fertile landscape has emerged in periodically driven systems \cite{zaletel_colloquium_2023}. In this case, the many-body interacting system is driven by a periodic (Floquet) Hamiltonian of period $T$, $\hat H(t) = \hat H(t+T)$. A Floquet time crystal (FTC) manifests itself in the response of a macroscopic observable exhibiting oscillations with a periodicity that is a multiple of the imposed drive, often observed as a period doubling at $2T$, but also generalized to higher orders \cite{surace_floquet_2019,giergiel_time_2018,giergiel_creating_2020,pizzi_higher-order_2021,giachetti_fractal_2023}. These anomalous oscillations exhibit enduring persistence as the system approaches the thermodynamic limit.
Therefore, a genuine breaking of time translation symmetry materializes only in this limit, akin to conventional spontaneous symmetry-breaking phenomena. Nevertheless, there is also the possibility that nonequilibrium phases break the time translation symmetry up to times that can be exponentially long with the tuning of the system parameters - the so-called prethermal regime - even though they do not diverge in the thermodynamic limit. In these cases, the system supports a  \textit{prethermal} Floquet time crystal (pFTC) \cite{else_prethermal_2017}.

 In general, a FTC (or pFTC) requires a mechanism in order to prevent thermalization, i.e., from heating up due to the Floquet dynamics. Various forms of ergodicity breaking mechanisms have been exploited in this context, such as those induced by disorder in many-body localization (MBL) \cite{else_floquet_2016,von_keyserlingk_phase_2016,khemani_phase_2016,yao_discrete_2017}, long-range interactions \cite{russomanno_floquet_2017,surace_floquet_2019,yang_dynamical_2021,morita_collective_2006,ojeda_collado_emergent_2021,nurwantoro_discrete_2019,pizzi_higher-order_2021,munoz_arias_floquet_2022,giachetti_fractal_2023}, Stark effect \cite{liu_discrete_2023}, many-body Scars \cite{maskara_discrete_2021,huang_analytical_2023}, high driving frequencies~\cite{else_prethermal_2017,kyprianidis_observation_2021,beatrez_critical_2023,pizzi2021classical,ye2021floquet,pizzi2021classicalPRB}, among others \cite{kyprianidis_observation_2021,stasiuk_observation_2023,PhysRevLett.130.180401,beatrez_critical_2023,makinen_magnon_2023,euler_metronome_2024}. 
Prethermal FTCs have been discussed in systems with disordered local external magnetic fields \cite{ippoliti2021many} (focus of this work), inducing a many-body lifetime increasingly  long with the disorder strength; or in systems driven by high-frequencies, thereby showing a lifetime that is exponentially long in the ratio of the system couplings to the drive frequency \cite{else_prethermal_2017}.  In this latter case, classical effective descriptions have been proposed able to capture the main phenomenology in the prethermal regime \cite{pizzi2021classical,ye2021floquet,pizzi2021classicalPRB}, indicating that such pFTCs may be independent (but robust) to quantum fluctuations. Interestingly, in this work we show that a quantum formalism becomes relevant in order to capture the capabilities of disordered pFTCs for quantum applications (in particular, in metrology protocols overcoming the standard quantum limit in precision).


Despite these varieties of mechanisms and their corresponding platforms, there are no studies of such phases in chiral multiferroic chains with their peculiar properties \cite{ramesh2007multiferroics,bibes2008towards,fiebig2005revival, cheong2007multiferroics,PhysRevLett.125.227201,PhysRevB.91.041408,khomeriki2016positive,wang2020optical,trybus2024dielectric}.
Chiral multiferroic systems are characterized by exchange interactions between nearest and next nearest neighbor spins ($J_{1,2}$). Besides, coupling to the electric field via the magneto-electric mechanism leads to the effective Dzyaloshinskii-Moriya interaction  (DMI), which is an extra source for controlling magnetic phases in the system. In addition, the chain can undergo an MBL regime in the presence of a random magnetic field \cite{stagraczynski2017many}. The level spacing statistics are influenced by the extended DMI interactions; while in the absence of DMI, one observes a transition from a Gaussian orthogonal ensemble to Poisson distributions by increasing the disorder strength, the presence of DMI introduces mixed symmetries with the appearance of Gaussian unitary ensemble distributions in the ergodic regime. These materials can be realized experimentally, offering great potential for quantum technologies due to their magnetic ferroelectricity \cite{cheong2007multiferroics}. One of the goals of this work is to exploit their properties in quantum sensing protocols. Concerning the experimental aspects of the fabrication of chiral spin chains, we refer to the work \cite{PhysRevLett.108.197204}. 

In fact, quantum systems offer great opportunities for sensing different types of fields due to their strong sensitivity to external perturbations, such as weak magnetic \cite{kominis2003subfemtotesla,budker2007optical,vengalattore2007high,taylor2008high}, electric \cite{kornack2005nuclear,brownnutt2015ion} and gravitational fields \cite{schnabel2010quantum}, with a wide range of applications from materials science \cite{lovchinsky2017magnetic} to medical testing \cite{jensen2016non}.
In addition to its high sensitivity to disturbances,
  the accuracy of quantum sensors can be enhanced by quantum correlation, such as entanglement~\cite{pezze_quantum_2018,PhysRevA.85.022321} or spin squeezing along preferred directions~\cite{block2024scalable,kitagawa1993squeezed}, allowing to overcome the limits for classical or non-interacting spin sensors -  also known as the Standard Quantum Limit (SQL) \cite{giovannetti_advances_2011}. 
The development of sensors that exploit the many-body nature of the system in a coherent form is generally not trivial. In particular, sensing AC fields becomes a challenging task. The most common approach in this case is the use of dynamic decoupling schemes, where high frequency spin echo pulses are employed to increase the coherent time of the sensor, thus allowing a long interrogation time with the estimated AC field \cite{hansom2014environment,muller2014optical,PhysRevLett.129.126101}. In such schemes, however, the required high-frequency pulses tend to mitigate the quantum correlations between the spins, thus do not exploiting the full potential of the sensor.
Recent approaches have been proposed in order to use spin interactions in a beneficial way for AC sensing rather than hindering it \cite{iemini2023floquet,mishra_integrable_2022,zhou2020quantum,choi_quantum_2017}. In particular, the use of time crystals for quantum applications has been a subject of growing interest recently  \cite{iemini2023floquet,gribben_quantum_2024,yousefjani_discrete_2024,bomantara_simulation_2018,montenegro_quantum_2023,lyu_eternal_2020,cabot_continuous_2023,carollo_nonequilibrium_2020,carollo_quantum_2023,bao_schrodinger_2024}, and their use as quantum sensors of AC fields is emerging as a promising avenue \cite{iemini2023floquet,gribben_quantum_2024,moon2024discretetimecrystalsensing}.

In this manuscript, our primary goal is to investigate the emergence of FTC in a chiral multiferroic chain once subjected to a periodic drive. Subsequently, we aim to explore their sensing capabilities in the detection of AC fields. We first show that pFTCs emerge in the limit of large disorder strengths, and derive the phase diagram of the model characterizing the dynamics of some main figures of merit, namely: the magnetization, coherence, and entanglement. Thereafter, we introduce an external AC field into the system and analyze its sensing properties in both the pFTC and ergodic phases. We tune the sensor parameters according to recent experimental realizations of chiral multiferroic chains in order to be closer to the realistic values. The sensor performance is quantified by the quantum Fisher information measure \cite{guo2017quantifying,gefen2017control,wang2022quantum, iemini2023floquet}. We observe an enhanced performance of the pFTC sensor once tuned in period doubling frequency to the AC field frequency.  Within the prethermal regime, the sensor overcomes the SQL and can reach a QFI scaling superlinearly with the number of spins.

The manuscript is organized as follows. In Sec. \ref{model}, we describe the chiral multiferroic chain and its Floquet dynamics, along with its experimental implementation. In Sec. \ref{time_crystal}, we explore the dynamical properties of the system and discuss the phase diagram of the model. In Sec. \ref{sensor_properties} we focus on the sensing capabilities of the system to external AC fields. Finally, in Section \ref{conclusion} we draw our main conclusions and perspectives.

\section{Model}
\label{model}
A chiral multiferroic chain can be described by the spin-1/2 Heisenberg model with nearest and next-nearest-neighbor interactions, along with an external random magnetic field and magnetoelectric coupling induced by an electric field:
\begin{eqnarray}
\label{Hamiltonian}
\hat{H}_0&&=J_1\sum_{i=1}^{N-1}{\bf \hat  S_i . \hat S_{i+1}}+J_2\sum_{i=1}^{N-2}{\bf \hat  S_i . \hat S_{i+2}} \nonumber \\
&&- \sum_{i=1}^Nh_i^z S_i^z+ D\sum_{i=1}^N{\bf (\hat  S_i \times \hat S_{i+1}})_z.
\end{eqnarray}
Here, the first two terms $J_1$ and $J_2$ describe Heisenberg exchange interactions, with  $ {\bf \hat S_i} = (\hat \sigma_x,\hat \sigma_y,\hat \sigma_z)/2$ representing the spin-1/2 operator at site $i$.  The third term involves an external random magnetic field $h_i^z$ uniformly chosen in the range $h_i^z \in [-h, h]$. The fourth term, attributed to an external electric field $E_y$ and magnetoelectric coupling $g_{\rm ME}$, encapsulates the Dzyaloshinskii-Moriya interaction $D=E_yg_{\rm ME}$ between nearest-neighbour spins. 

This last term is a material specific for chiral multiferroics with broken inversion symmetry, and it plays an essential conceptual role in forming the system's ground state, influencing the transition between MBL and ergodic phases, among other features \cite{stagraczynski2017many,sinner2024superconducting,vijayan2023topological,cheong2007multiferroics,menzel2012information,mostovoy2006ferroelectricity,katsura2005spin,park2007ferroelectriccity,schrettle2008switching,shukla2023quantum}.
In fact, the DMI is the essence of the magneto-electric coupling between ferroelectric polarization and the external electric field, allowing the system's dynamics to be controlled not only by an external magnetic field but also by the electric field.

For the investigation of FTC properties in the chiral multiferroic chain, we consider a kicking operator $\sum_i \hat S_i^x$ in the system inducing a rotation of $\phi$ in the spins around the x-axis. Consequently, our Hamiltonian takes the form:
\begin{equation}
  \hat H(t) =\hat H_0 +\sum_{n=0}^{\infty}\delta(t-nT) \phi\sum_{i=1}^N  \hat S_i^x .
  \label{HF}
\end{equation}
The Floquet unitary for the dynamics within a single period $T$ is given by,
\begin{equation}
   \hat  U= \hat U_{\rm kick}\exp(-i\hat H_0T),\,\, {\rm with} \,\, \hat U_{\rm kick}=\exp\Big[-i\phi \sum_{i=1}^N\hat S_i^x\Big]
\end{equation}
where for simplicity, we consider $\hslash=1$.

\begin{figure*}
\includegraphics[width=.3\linewidth, height=.24\linewidth]{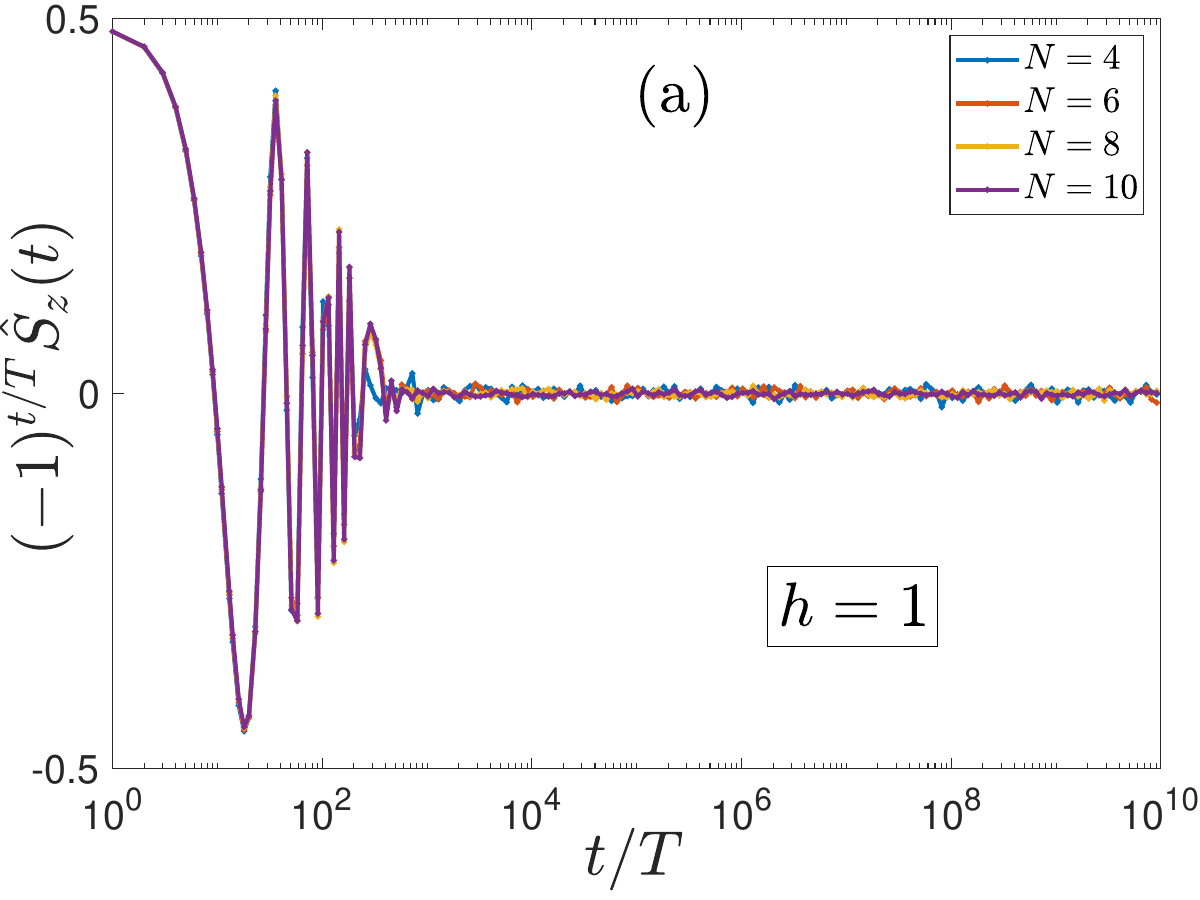}
\includegraphics[width=.3\linewidth, height=.24\linewidth]{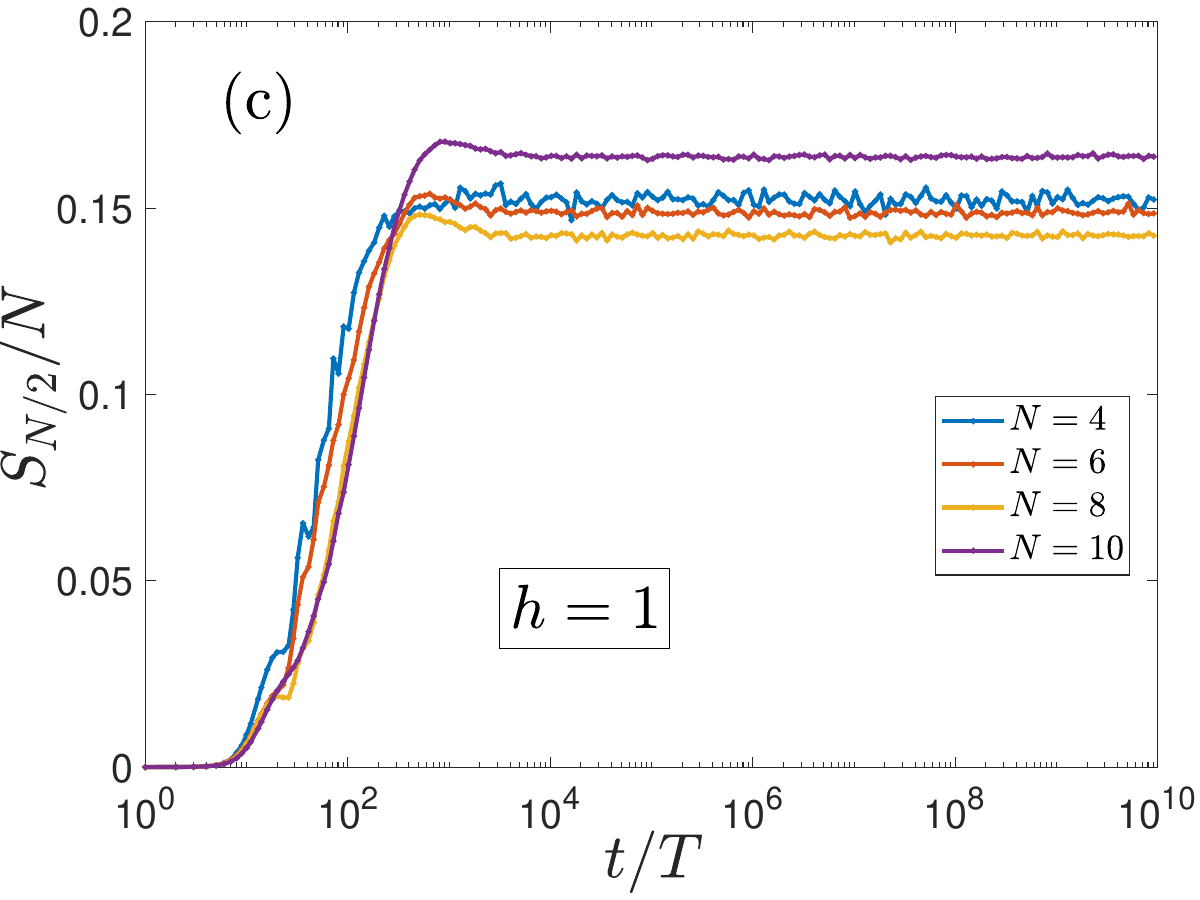}
\includegraphics[width=.3\linewidth, height=.24\linewidth]{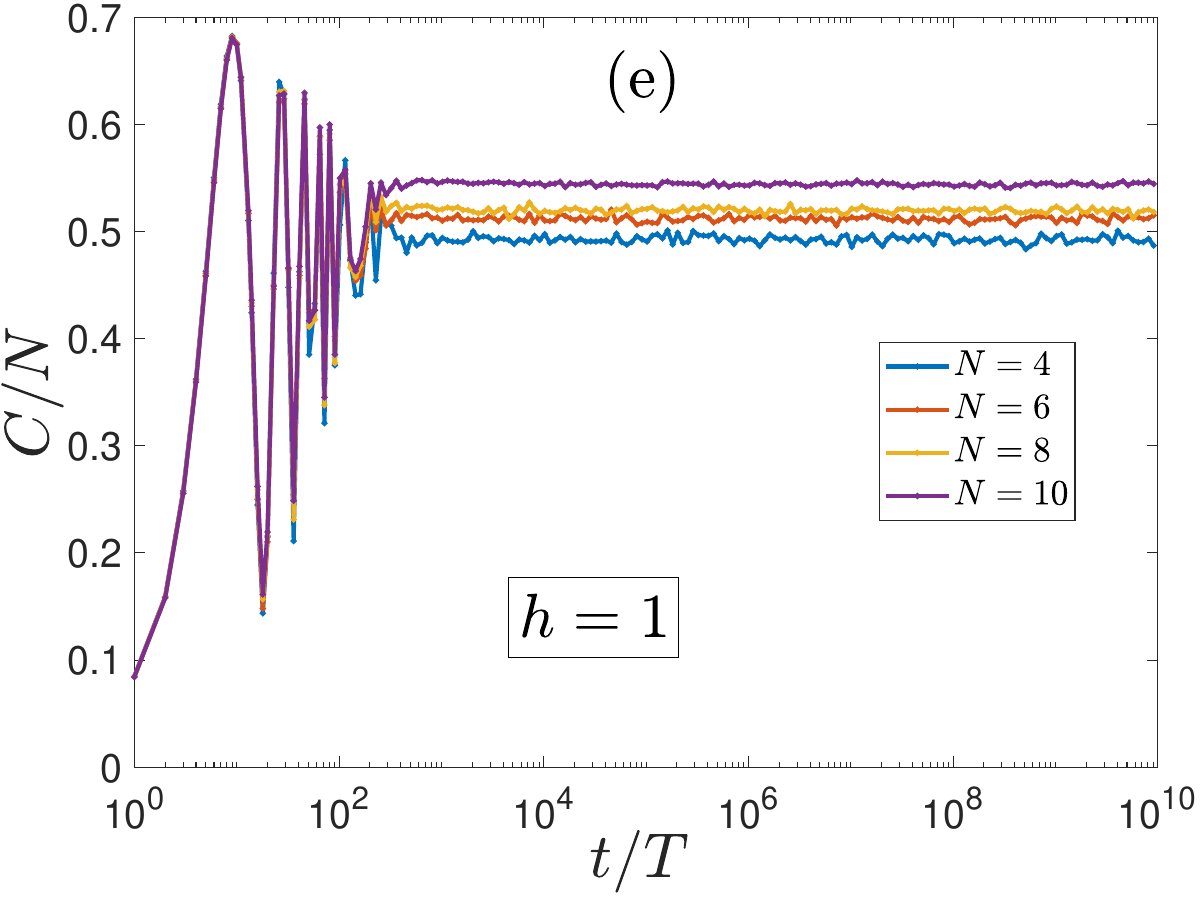}
\includegraphics[width=.3\linewidth, height=.24\linewidth]{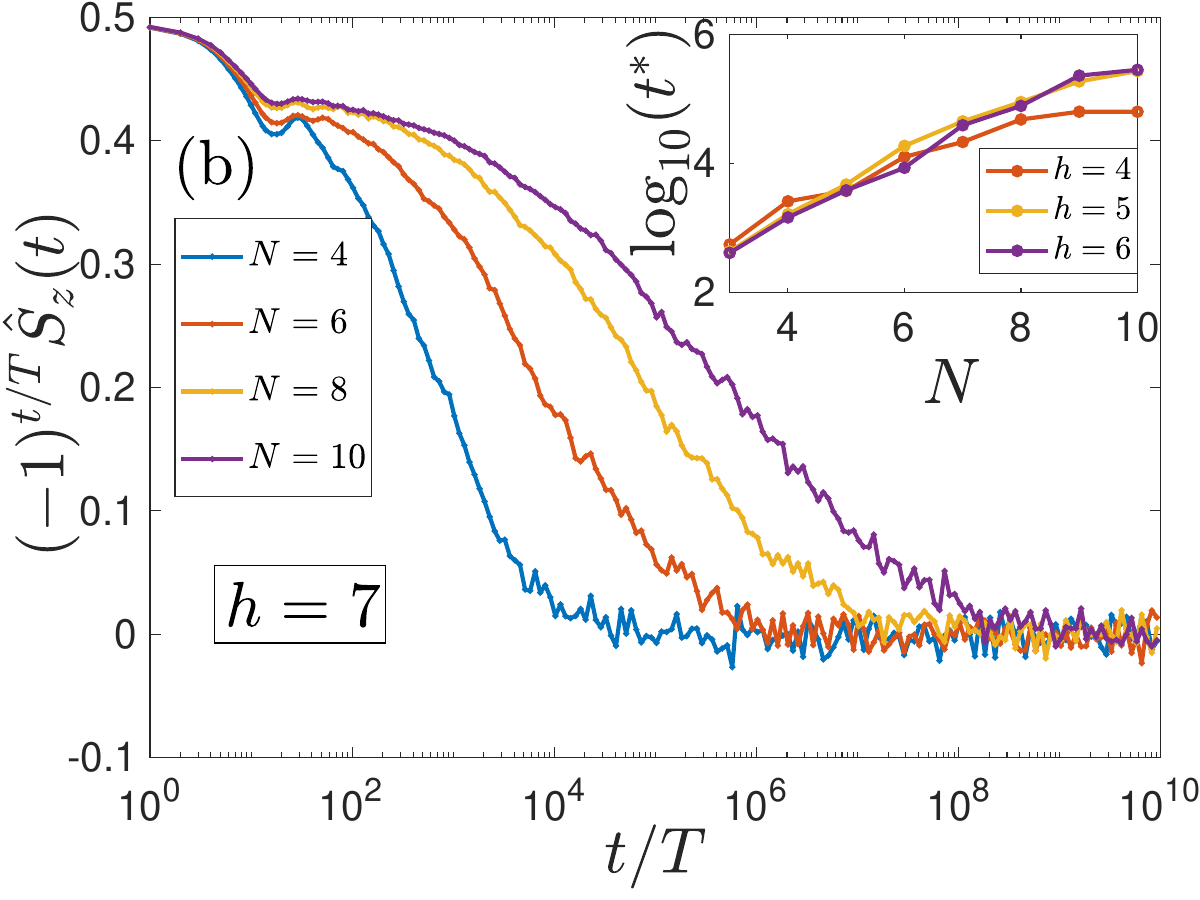}
\includegraphics[width=.3\linewidth, height=.24\linewidth]{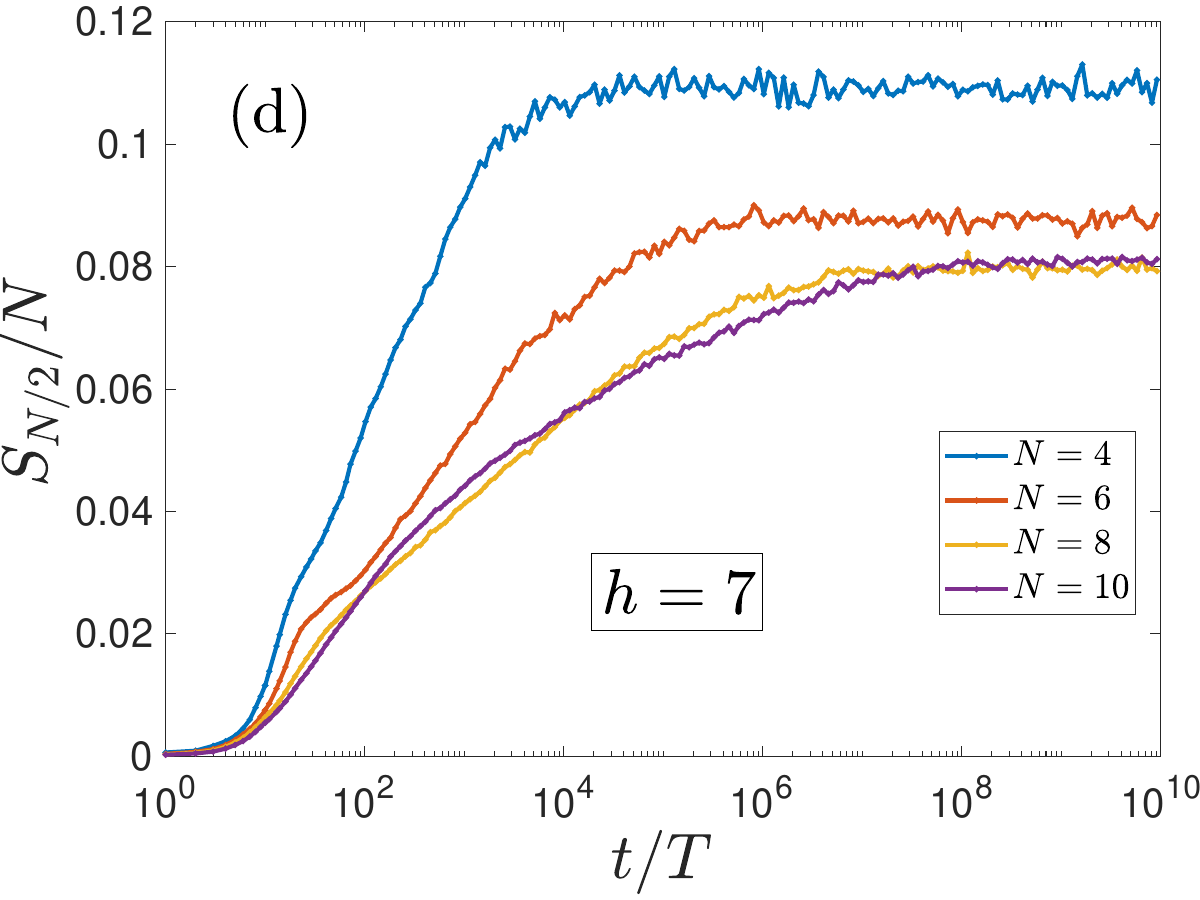}
\includegraphics[width=.3\linewidth, height=.24\linewidth]{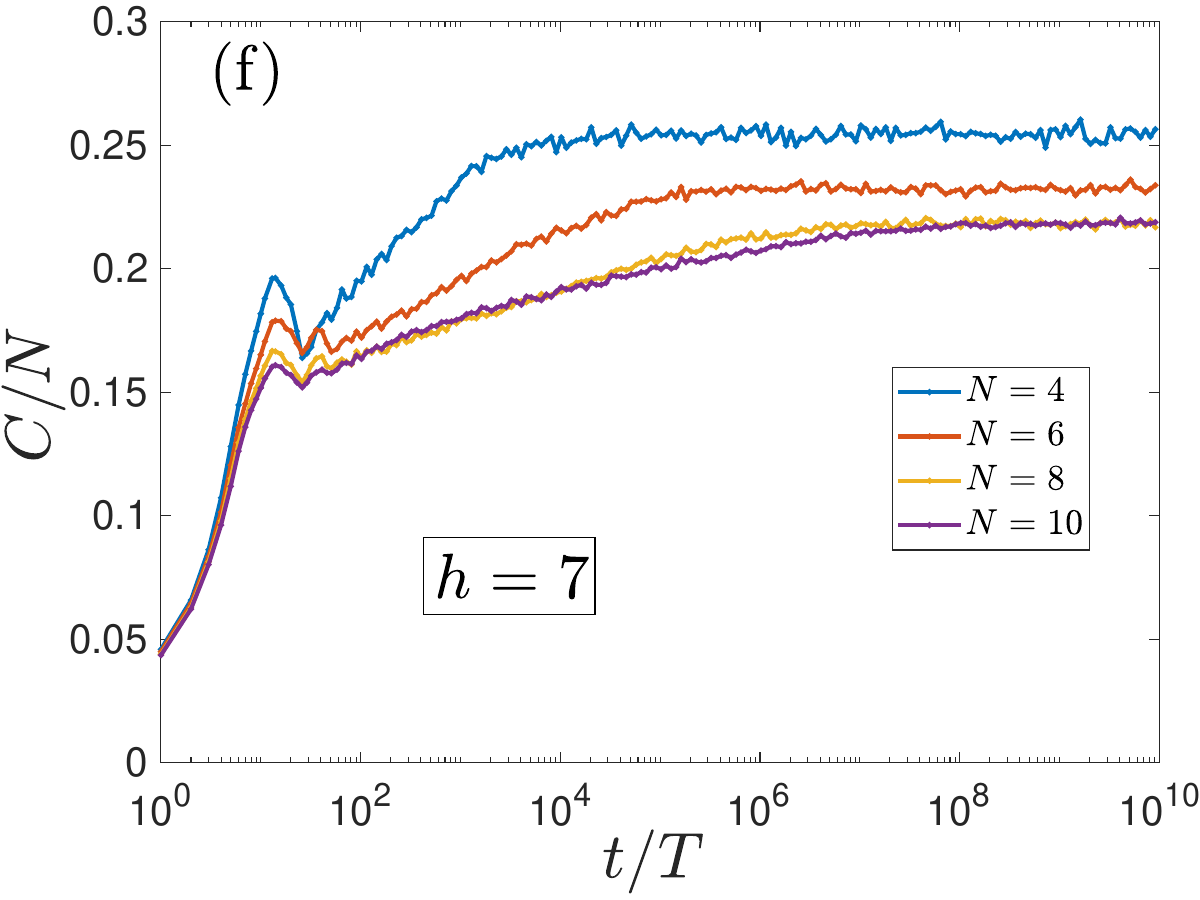}
\caption{ Magnetization, bipartite entanglement entropy, and coherence as a function of time for small disorder parameter $h=1$ in (a), (c), and (e), and for large disorder parameter $h=7$ in (b), (d), and (f), respectively. The parameters are $J_1 = -1$, $J_2 = -J_1/4$, $D = 0$, and $\phi = 3.05$. A total of $2 \times 10^3$ disorder realizations were taken into account.  The inset in panel (b) shows the lifetime as a function of the sensor size for various values of $h$. The parameters are set to $J_1 = -1$, $J_2 = -J_1/4$, $D = 0$, and $\phi = 3.0$. The data represents an average over 500 disorder realizations. } 
   \label{Magnetization_linear}
  \end{figure*}

\subsection*{Experimental feasibility}

In our numerical simulations of the model we set the theoretical parameters close to those from recent experimental realizations of chiral multiferroic chains~\cite{yasui2013dielectric,masuda2005spin,solodovnikov1997new,trybus2024dielectric}. 
In particular, we focus on polycrystalline samples of $\rm Rb_2Cu_2Mo_3O_{12}$. 
 The crystal structure of such materials consists of one-dimensional chains of $\rm Cu^{2+}$ ions aligned along a specific axis, with $\rm MoO_4$ tetrahedra and $\rm Rb^+$ ions positioned between these chains. 
 The $\rm Cu^{2+}$ ions are linked by oxygen ions forming twisted $\rm CuO_{2}$ ribbon chains.  The $ \rm Cu^{2+}$ ions form in this way quasi-one-dimensional spin chains with effective spin $S=1/2$. The exchange interactions between these $\rm Cu^{2+}$ spins are ferromagnetic ($J_1=-138 {\rm K}$) and antiferromagnetic ($J_2=51 {\rm K}$).  DMI arises due to magneto-electric coupling, with $g_{\rm ME}$ at the order of $0.1J$~\cite{trybus2024dielectric}. 
 Therefore, in our analysis, we consider Heisenberg exchange couplings at the order of $\vert J_2/J_1 \vert \approx 1/4$, and $D$ in the range $0.01\leq D/J \leq 1$ as realistic values. Moreover, magnetic fields $h$ are external parameters that can be tuned in these systems as e.g. using  magnetic property measurement system (MPMS), which applies a controlled magnetic field in the material via superconducting solenoids~\cite{yasui2013magnetic}, among other techniques~\cite{menzel2012information,park2007ferroelectriccity,schrettle2008switching,mostovoy2006ferroelectricity,katsura2005spin} .

\section{Prethermal Floquet Time Crystal }
\label{time_crystal}

In this section, we study the emergence of pFTCs in the model, discussing the behavior of its dynamical properties - magnetization, coherence, and entanglement - and their dependence on the system parameters. 
 These properties have unique behaviors in pFTCs, with their ability to exhibit   increasingly  slow heating while sustaining quantum coherent and correlated dynamics. Therefore understanding the dependence of these properties with the tuning of system parameters shall be directly relevant for quantum applications of such phases as sensors.

  \subsection*{Finite size scaling}
  
The effects of disorder in chiral multiferroic chain were examined in Ref.~\cite{stagraczynski2017many}, revealing distinct phases depending on the disorder strength: for small values of $h$, an ergodic phase was observed, while increasing the disorder strength led to a transition into the many-body localization (MBL) phase. We first analyse the period doubling magnetization dynamics in both of these regimes, highlighting the emergence of pFTC for large enough disorder strength and system sizes. Precisely, we compute the magnetization, 
\begin{equation} \langle \hat{S}_z(t) \rangle = \frac{1}{N} \bra{\psi(t)}  \left(\sum_{i=1}^N \hat{S}_i^z \right) \ket{ \psi(t)}, 
\end{equation}
where $\ket{\psi(0)}$ is a separable initial state given by \begin{equation} \label{Initial_state} 
\ket{\psi(0)} = \otimes_{j=1}^N \left[ \cos\left(\frac{\theta_j}{2}\right) \ket{\uparrow} +   \sin\left(\frac{\theta_j}{2}\right) \ket{\downarrow} \right], 
\end{equation} with $\theta_j = \pi/16$ . Here, $\ket{\downarrow}$ and $\ket{\uparrow}$ are the eigenstates of the Pauli operator $\hat{\sigma}_z$. 
 We show in Fig.~\ref{Magnetization_linear}  our results for a kicking phase $\phi  \approx \pi$. Due to the periodic nature of the dynamics, the observables are displayed at stroboscopic times, i.e., at integer multiples of $t/T$. For a low disorder strength $h = 1$, the collective magnetization dephases with amplitude decreasing rapidly until saturation at a null value; the behavior is moreover roughly independent of the system size - see Fig.~\ref{Magnetization_linear}(a). 
 On the other hand, for a large disorder strength $h = 7$, the magnetization displays a period doubling dynamics which remains finite for a significantly longer time. Notably, the time required for the magnetization to reach zero increases with system size but does not diverge in the thermodynamic limit due to its prethermal behavior - see Fig.~\ref{Magnetization_linear}(b).  Note that the pFTC behavior in our system originates from the disordered term in the external field, while the coupling (interaction) terms remain unaffected.
 In this way, although disorder may lead to a slower dynamics, its effective strength is greatly reduced by the kicking in the Floquet evolution. Precisely, at even stroboscopically times the effective kicking tends to counteract the local magnetic fields (as a destructive echoing effect) thus mitigating the disordered potential. A detailed discussion is provided in Ref. \cite{ippoliti2021many}. 
  We perform a finite-size scaling analysis of the lifetime, {\it i.e.}, $\log(t^{*})$ as a function of the sensor size $N$ [inset of Fig.~\ref{Magnetization_linear}(b)]. The results show that the lifetime initially grows exponentially with the sensor size and eventually saturates to a finite value in the thermodynamic limit.
 
In order to investigate the entanglement dynamics in these regimes~\cite{sahu2024}, we compute the entanglement entropy of the system, i.e., the von Neumann entropy of the reduced density matrix $\hat{\rho}_X$,
\begin{equation}
S_X = \text{Tr}\left[-\hat{\rho}_X \log_2 \hat{\rho}_X \right], 
\end{equation} 
where $\hat{\rho}_X = \text{Tr}_{X^c} [\hat{\rho}]$ is obtained by tracing out the complementary subsystem $X^c$ from the full density matrix $\hat{\rho} = \ket{\psi(t)}\bra{\psi(t)}$.  In our analysis, we consider the entanglement entropy for an equal bipartition and denote it as $S_{\frac{N}{2}}$.
In the ergodic phase (low disorder strength), the entanglement increases, eventually saturating at a volume-law constant value - Fig.~\ref{Magnetization_linear}(c). The time required for the entanglement to saturate corresponds roughly to the time at which the magnetization reaches zero. In the pFTC phase, similar to the ergodic phase, the entanglement also grows till a saturation value which follows a rough volume-law scaling, with a time required matching the one for which the magnetization approaches zero [Fig.~\ref{Magnetization_linear}(d)]. We notice, however, that the value of saturation is significantly lower than in the ergodic case.

This behavior trend is also observed along the coherence dynamics, which embodies both the essence of entanglement but also quantifying the degree of superposition of a quantum state with respect to a particular reference basis. 
 In our system a natural reference basis to be considered follows by the spin magnetisation along the z-direction, due to its peculiar stability properties along the pFTC phase. There are different forms to quantify the coherence in a quantum state  \cite{streltsov2017colloquium}, we use here the one based on \textit{relative entropy of coherence} defined as:
\begin{equation}
C(\hat \rho) = S(\hat \rho_{\text{diag}}) - S(\hat \rho),
\end{equation}
where $S(\hat \rho)$ is the von Neumann entropy of the state $\hat \rho=\ket{\psi(t)}\bra{\psi(t)}$, and $\hat \rho_{\text{diag}}$ is the diagonal part of $\hat \rho$ in the reference basis.
We obtained, similar to entanglement dynamics, that for both low and higher disorder strengths, the dynamics of coherence increase till a saturation which is roughly independent of system size for the low disorder while displaying finite size corrections for higher disorders - see Fig.~\ref{Magnetization_linear}(e-f)]. In both cases, the time required for the coherence to saturate is approximately equal to the magnetization, attaining a null value, and higher disorder strengths tend to saturate at a lower coherence value.

\begin{figure*}
\includegraphics[width=.32\linewidth, height=.22\linewidth]
{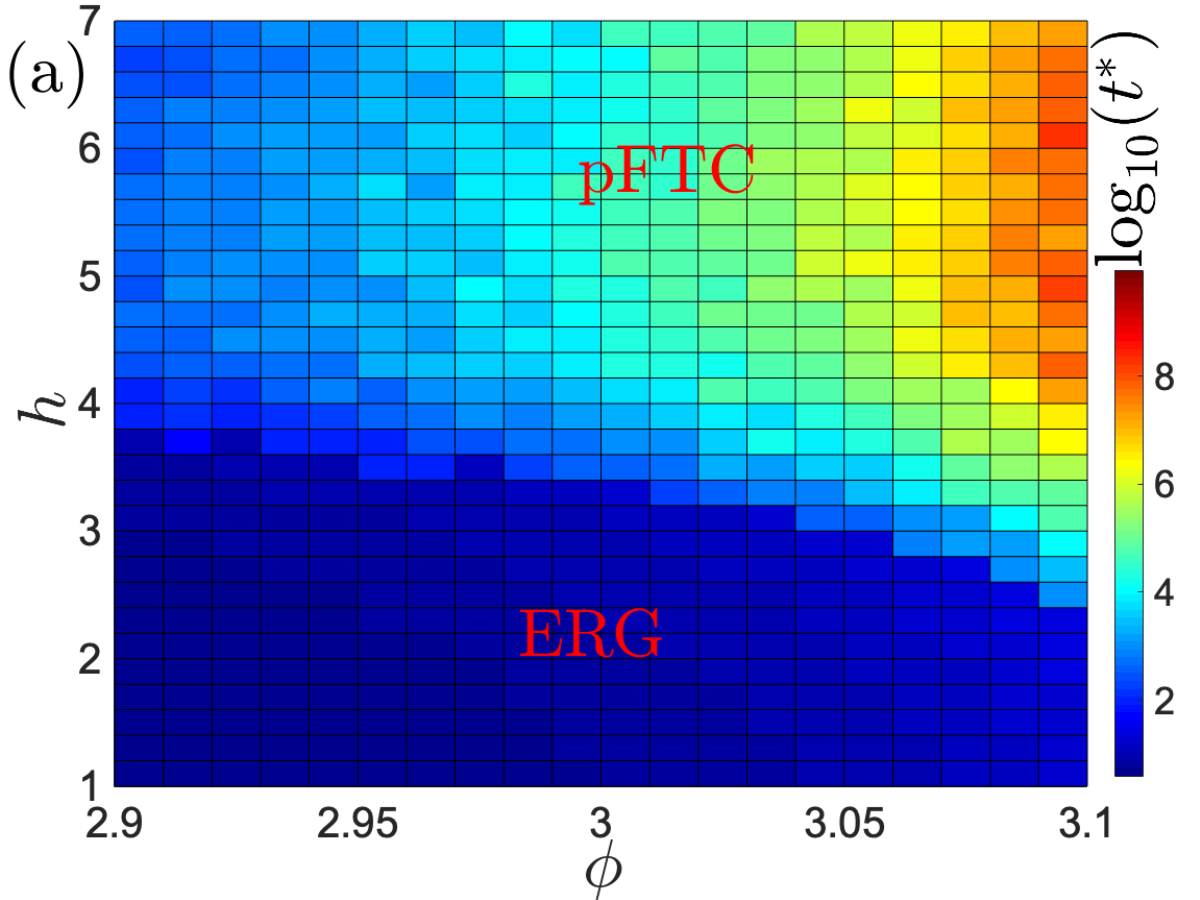}
\includegraphics[width=.32\linewidth, height=.22\linewidth]
{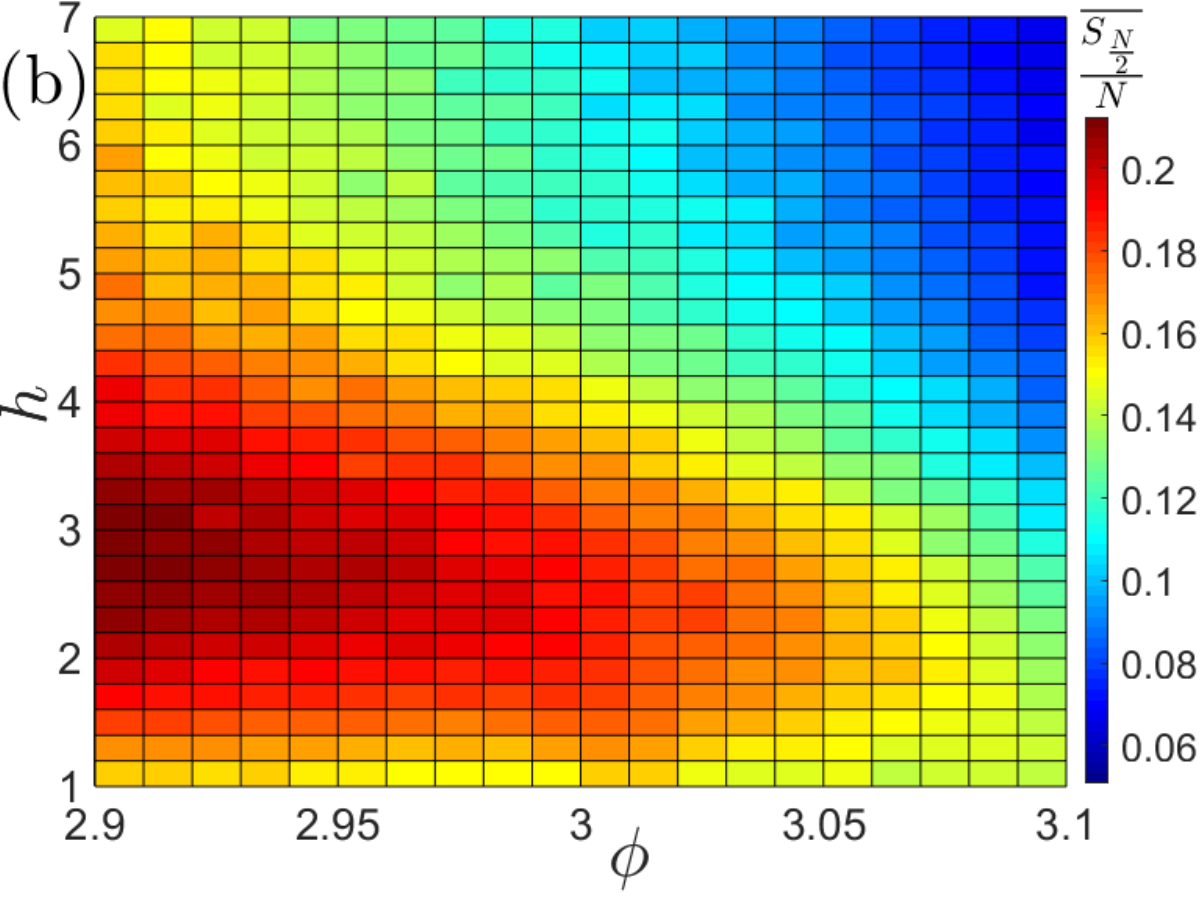}
\includegraphics[width=.32\linewidth, height=.22\linewidth]
{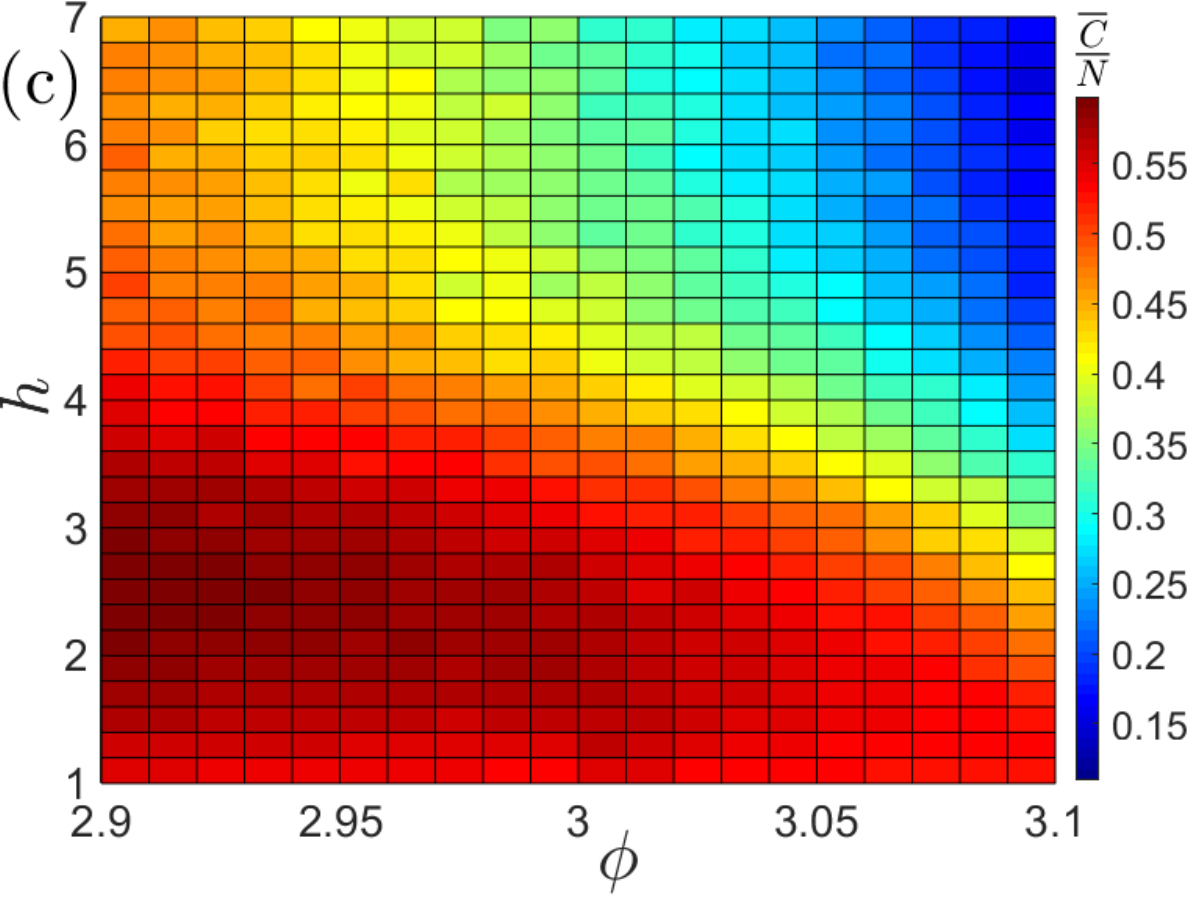}
\caption{(a) Lifetime of the pFTC, (b) saturation value of entanglement and (c) saturation value of coherence with disorder parameter $h$ vs $\phi$.  Parameters are $J_1=-1$, $J_2=-J_1/4$, $D=0$, $N=8$. $250$ disorder realization taken into account.} 
   \label{mag_h_phi}
  \end{figure*}

\subsection*{Phase diagram}

One of the central features of pFTC's, namely the    increasingly  slow heating, arises due to the disorder in the model Hamiltonian. 
Since the disorder in the Hamiltonian is only along the external field and not among the coupling (interacting) parameters, the kicking unitaries tend to cancel the effect of the external field disorder \cite{ippoliti2021many}, undermining the many-body localization in the system. 
Therefore the system shall always thermalize at a finite time (lifetime) independent of the number of spins in the chain. 
Formally, we define the lifetime of the system dynamics as the time it takes for the magnetization to approach a negligible value ($t^*$ such that $\vert \hat S_z(t^*) \lesssim  \epsilon$, where $\epsilon = 10^{-2}$).
  Our results reveal that the pFTC lifetime depends distinctly on the kicking phase $\phi$ and the strength of the disordered magnetic field $h$. Specifically, the lifetime increases as one approaches the perfect $\pi$-kick phase, following a power-law scaling: $t^{*} \sim \vert \pi - \phi \vert^{-\gamma}$, where $\gamma > 0$. The dependence on the disorder strength $h$ exhibits different scaling behaviors. In the ergodic phase, the lifetime remains approximately independent of the disorder strength. However, upon entering the pFTC regime, the lifetime increases rapidly with $h$, approximately following an exponential growth, $t \sim e^{\beta h}$ with $\beta >0$. For even larger disorder strengths, the scaling behavior changes, with the lifetime increasing as a power-law with $h$. For a detailed analysis, we refer to Appendix \ref{appendix}.

In Fig.~\ref{mag_h_phi}(a) we show our results for the dependence of the lifetime with the system parameters $h$ and $\phi$. We observe that when $\phi$ deviates significantly from the exact spin flip $\phi=\pi$ or decreases the disorder strength, the lifetime becomes shorter. In this region of low disorder or large deviation from the exact spin flip the lifetime is at the order of $O(10^1)-O(10^2)$ stroboscopic times and roughly independent of the system parameters; while otherwise, it can be significantly increased up to the order to $O(10^9)$ stroboscopic time for the considered system size. 
 Therefore, despite the always thermalizing nature of the model, the timescale of its thermalization is highly tunable through control of disorder and spin-flip precision, featuring moreover a high sensitivity to them as observed from sharp drops on its lifetime when the system moves away from ideal conditions. 
This sharp transition allows us to draw a phase diagram for the model, delineating the regions for its ergodic and pFTC phases.

We also discuss the dependence of the entanglement and coherence saturation values in comparison to the dynamics lifetime.  The average saturation value for the system observables is defined as,
\begin{eqnarray}
    \overline{O} = \frac{1}{T_2 - T_1} \sum_{t/T = T_1}^{T_2} O(t),
    \label{avg_Ent}
\end{eqnarray}
where $T_1$ and $T_2$ are the initial and final times, respectively, and $O$ is the observable of interest. We define the average times $T_{1(2)}$ in the saturation region, where $T_1$ represents the approximate initial time when the observable $O$ begins to saturate, and $T_2$ a larger time in order to obtain an accurate average value. 

The results show an interesting trade-off between the stability of the pFTC phase and the level of quantumness in the saturated system in terms of entanglement and coherence - see [Fig.~\ref{mag_h_phi}(b,c)]. 
Longer pFTC lifetimes can be associated with shorter correlation lengths in the underlying MBL localized states. Therefore, correlations shall propagate slowly in this regime. However, despite its velocity, after a long saturation time, the entanglement has spread along the whole spin chain, and the system always thermalizes. We find that despite this full thermalization, the trade-off between long pFTC lifetimes (short localization length) and lower quantumness is maintained even for the saturated system. Analogously shorter pFTC lifetimes allow for higher entanglement and coherence in the saturated system.
This has important implications for quantum system design. Depending on the desired outcome, one might prioritize either extending the pFTC phase for greater stability or enhancing its quantum correlations for applications. The nuanced relationship between these properties offers, in this way, both a deeper understanding of the dynamics of such non-equilibrium quantum systems as well as opportunities for optimizing them on specific quantum tasks.
\begin{figure*}
\includegraphics[width=.32\linewidth, height=.22\linewidth]{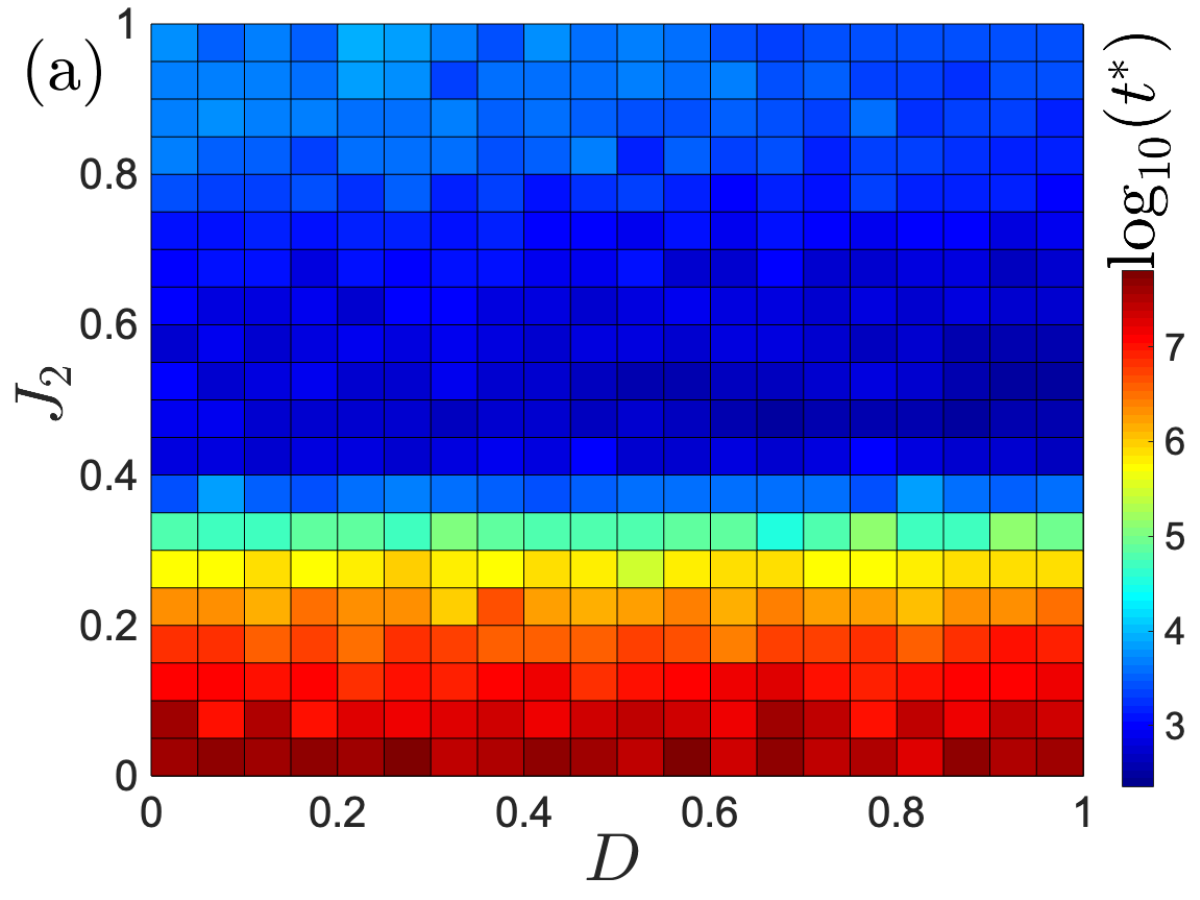}
\includegraphics[width=.32\linewidth, height=.22\linewidth]{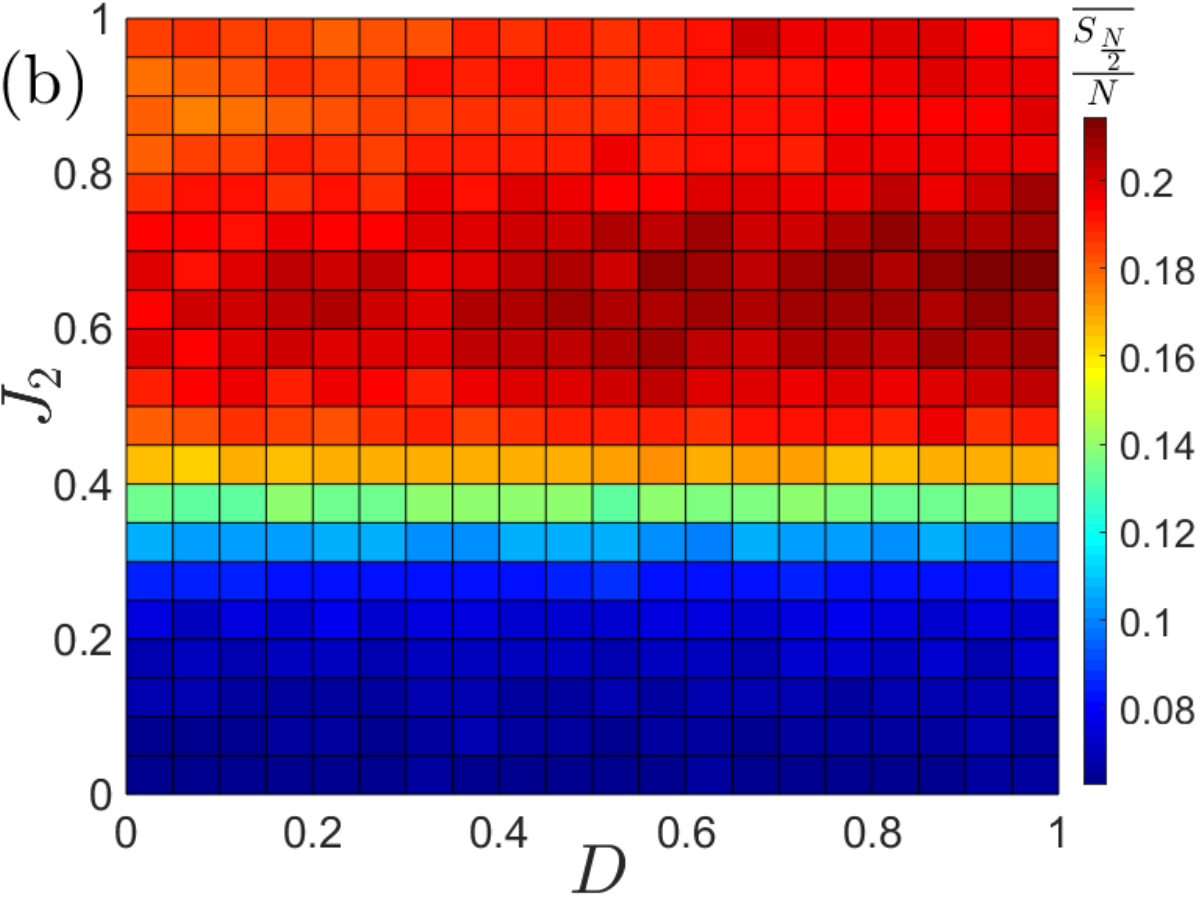}
\includegraphics[width=.32\linewidth, height=.22\linewidth]{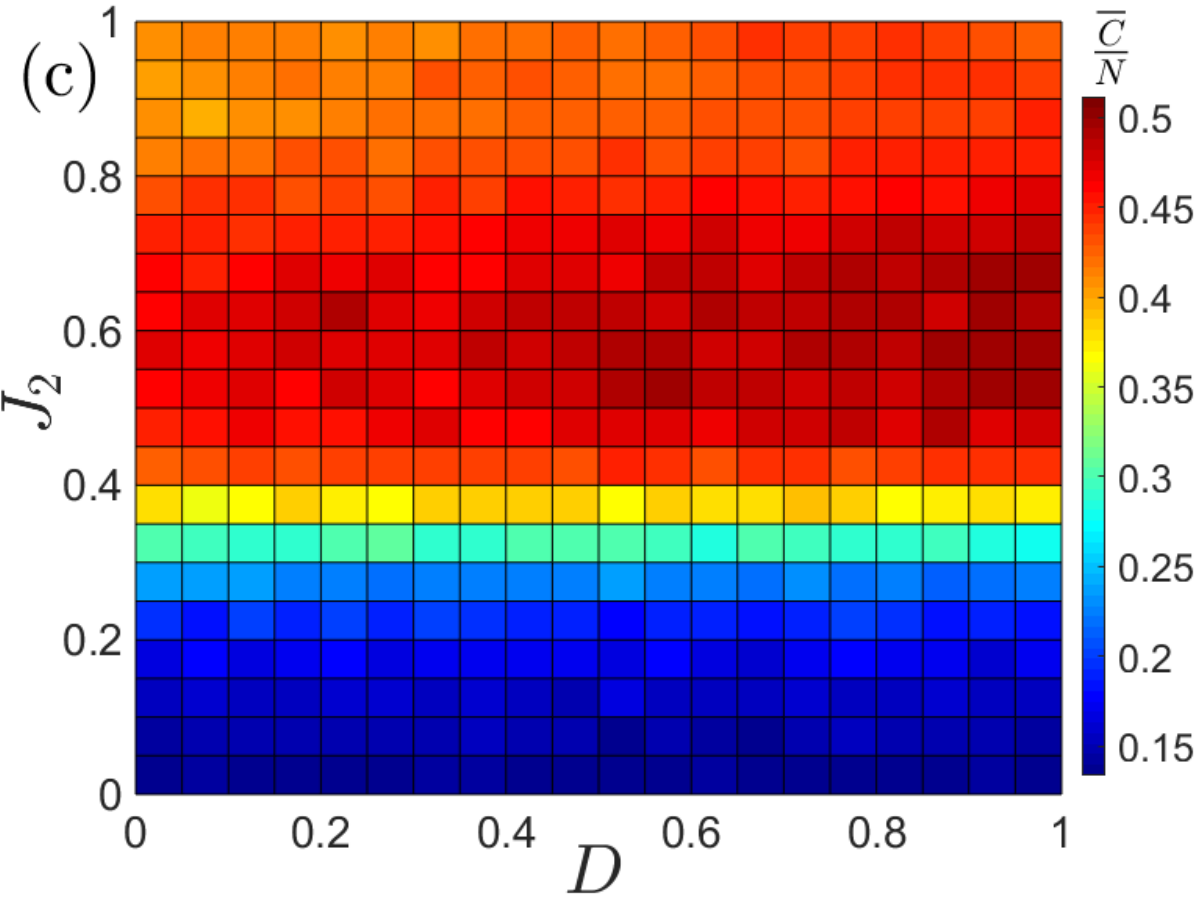}
\caption{(a) Lifetime of the prethermal FTC (b) saturation value of entanglement (c) saturation value of coherence with $J_2$ vs $D$.  Parameters are $J_1=-1$,  $N=8$ and $250$ disorder realization is taken into account,  $\phi=3.05$ and $h=7$. } 
   \label{J2_D_tl}
  \end{figure*}

\begin{figure}
\includegraphics[width=.85\linewidth, height=.50\linewidth]{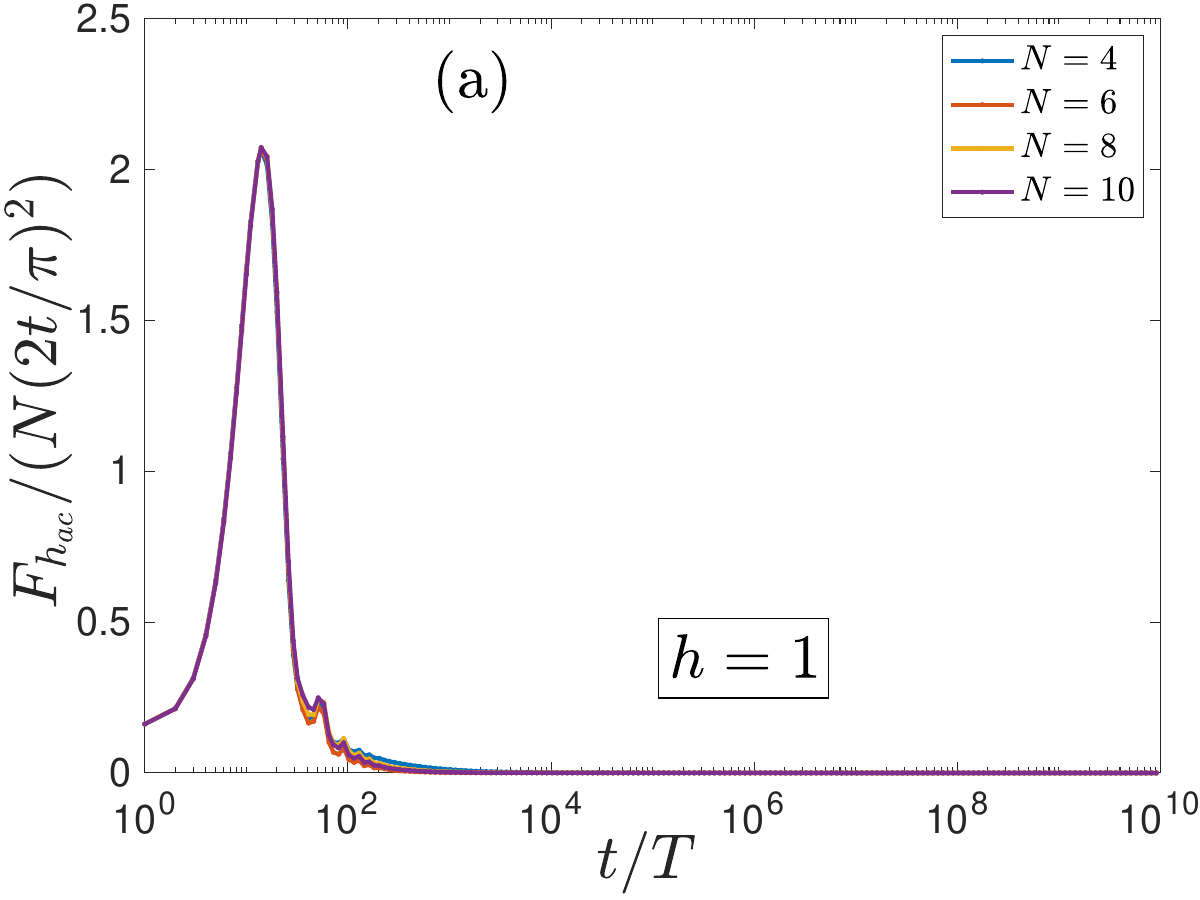}
\includegraphics[width=.85\linewidth, height=.50\linewidth]{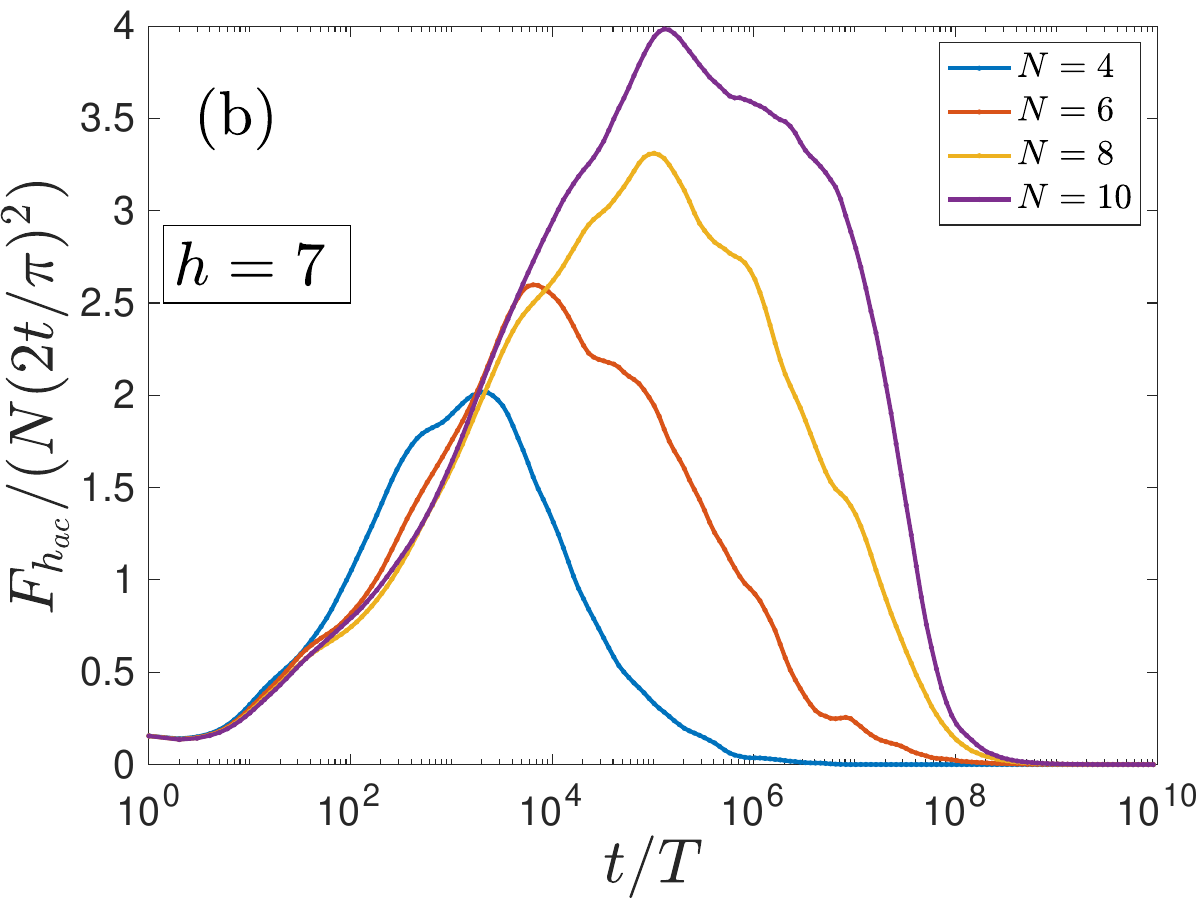}
\caption{(a) $F_{h_{ac}}/(N(2t/\pi)^2$) vs time $t$ with changing the size of the sensor in the (a) ergodic phase with $h=1$, and  (b) in the pFTC phase with $h=7$. Parameters used are \(J_1 = -1\), \(J_2 = -J_1/4\), \(D = 0\),  \(\phi=3.05\), with \( 10^3\) disorder realizations considered.} 
   \label{QFI_N}
  \end{figure}
\begin{figure*}
\includegraphics[width=.32\linewidth, height=.22\linewidth]{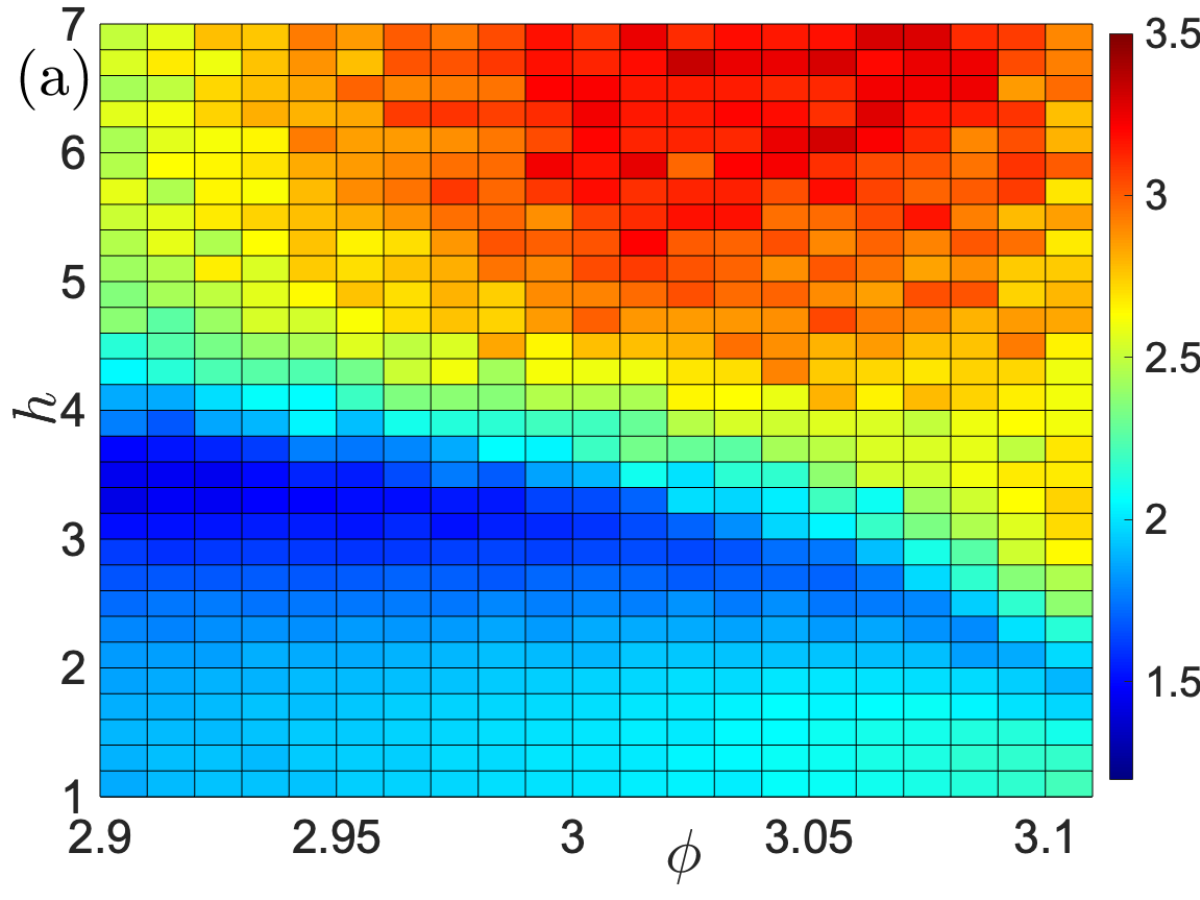}
\includegraphics[width=.32\linewidth, height=.22\linewidth]{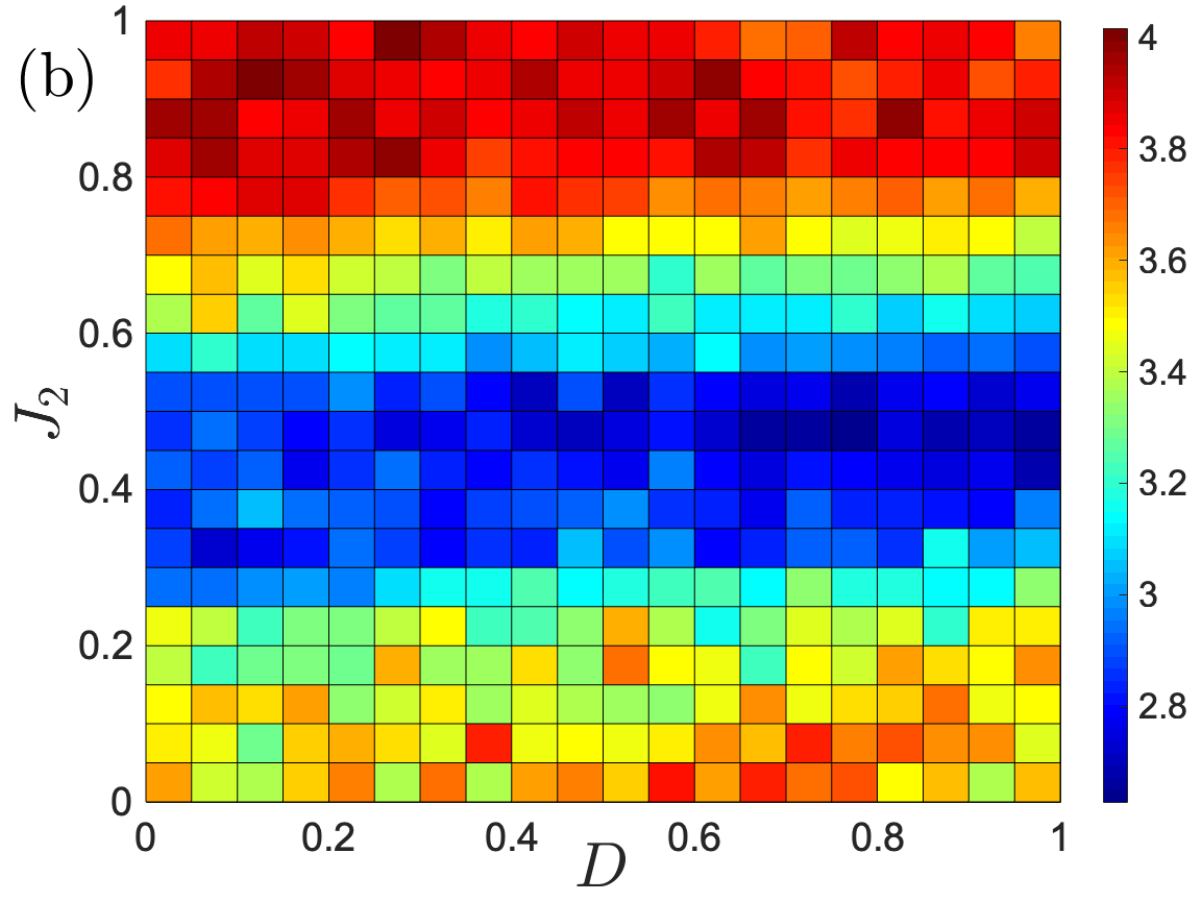}
\includegraphics[width=.32\linewidth, height=.22\linewidth]{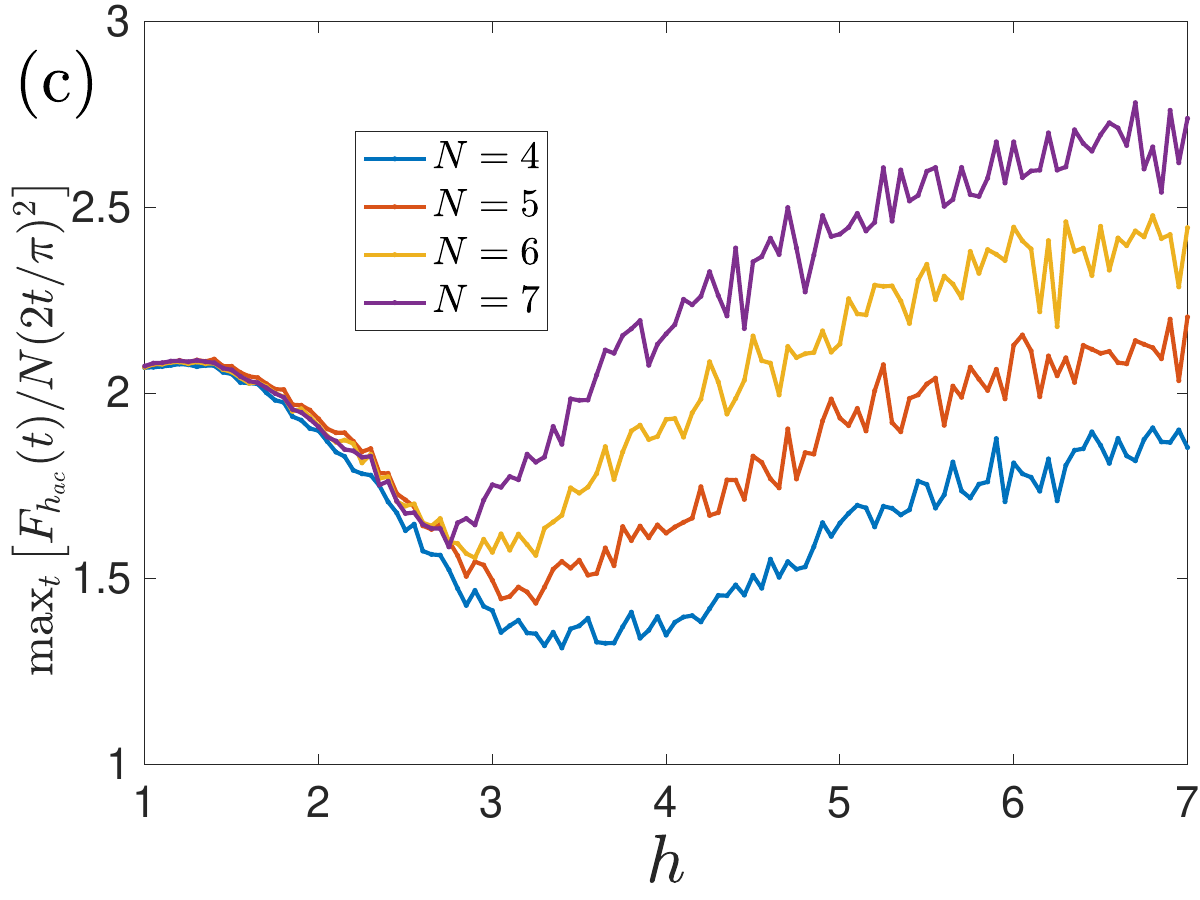}
\caption{ Maximum sensor performance with time, $ \max_t \left[ F_{h_{ac}}(t)/N(2t/\pi)^2 \right]$ as a function of (a) kicking angle \(\phi\) and varying disorder strength (parameters: are \( J_1 = -1 \), \( J_2 = -J_1/4 \), \( D = 0 \), and \( N = 8 \), with \( 10^3 \) disorder realizations taken into account); and (b) of the nnn $(J_2$) and DMI ($D$) interaction strengths (parameters:  \( J_1 = -1 \), \( \phi = 3.05 \), \( h = 7 \), and \( N = 8 \), with \( 10^3 \) disorder realizations taken into account). (c) Maximum QFI, $ \max_t \left[ F_{h_{ac}}(t)/N(2t/\pi)^2 \right]$, as a function of the disorder parameter \(h\), showing the impact of varying the size of the sensor \(N\) (parameters: \( J_1 = -1 \), \( J_2 = -J_1/4 \), \( D = 0 \), and \( \phi = 3.05 \), with \( 2 \times 10^3 \) disorder realizations considered).
} 
   \label{maxQFI}
  \end{figure*}
We also analyze the effect of nnn and DMI couplings on the lifetime and properties of the pFTC phase.
In chiral multiferroic systems, while nn couplings ($J_1$) are ferromagnetically ordered, the competing nnn ($J_2$) emerges antiferromagnetically. Therefore, this competing term favors conflicting spin alignments, which, in equilibrium situations, can result in frustrated spins and lead to non-collinear spin configurations such as spiral or disordered spin-glass phases \cite{PhysRevLett.96.067601}.
Our results show that introducing this ``frustrating" competition in the non-equilibrium Floquet dynamics tends to disrupt the pFTC phase. Specifically, when $\vert J_2/J_1 \vert\sim 1/2$  the system lifetime become minimal - see Fig.~\eqref{J2_D_tl}(a). Likewise, entropy and coherence saturation values exhibit larger saturation in the frustrated region - Fig.~\eqref{J2_D_tl}(b)-(c). 

On the other hand, DMI couplings introduce an antisymmetric exchange interaction between spins, leading to non-collinear equilibrium spin configurations. This interaction breaks inversion symmetry, favoring a preferred handedness in spin arrangements and potentially resulting in phenomena like skyrmions or chiral domain walls in the case of two-dimensional systems \cite{vijayan2023topological}. 
However, although DMI typically induces complex spin textures, its effect on the system's dynamics, including the lifetime of the pFTC, entanglement, and coherence, appears minimal within the explored parameter space ($0$ to $1$), with no observable dependence on DMI strength [Fig.~\ref{J2_D_tl}(a), (b), and (c)].

\section{Sensing of AC field} 
\label{sensor_properties}

  In this section, we discuss the potential of the Floquet chiral multiferroic chains as quantum sensors of an AC fields, represented by a sinusoidal wave with frequency $\omega_{\rm AC}$, phase $\theta_{\rm AC}$ and \textit{unknown} amplitude $h_{ac}$. Specifically, an AC field Hamiltonian expressed as,
\begin{equation}
   \hat H_{\rm AC}(t)= h_{ac}  \sin(\omega_{\rm AC} t + \theta_{\rm AC})\sum_{i=1}^N\hat S_i^z.
\end{equation}
  The full Hamiltonian for the spins in the multiferroic chain is therefore given by its internal one in addition to the unknown AC field term,
    \begin{eqnarray}
\hat H (t)= \hat H_0(t)+ \hat H_{\rm AC}(t), 
\end{eqnarray}
In order to exploit its sensing performance, we recall the standard procedure for the operation of a quantum sensor, involving a few basic steps:

(i) Initialization of the sensor, possibly in advantageous states leveraging quantum correlations among its constituents;

(ii) Interaction of the sensor with the signal of interest, allowing them to evolve together;

(iii) Performing a final measurement on the sensor state.

Given steps (i)-(ii), one can employ estimation methods on the final data in order to optimally extract the information about the encoded parameter $h_{ac}$ in the spins of the sensor. The minimum uncertainty in the estimated field ($\Delta h_{ac}$) is constrained by the Cramer-Rao bound,
\begin{equation}
\Delta h_{ac} = \frac{1}{\sqrt{F_{h_{\rm{ac}}}}}.
\end{equation}
where $F_{h_{\rm{ac}}}$ is the quantum Fisher information of the system \cite{liu2020quantum}, measuring the sensitivity of the sensor to the external unknown field. The quantum Fisher information is formally defined as,
\begin{equation}
\label{eq.qfi}
 \frac{F_{h_{\rm{ac}}}(t)}{4} = \langle \psi'(t,h_{\rm{ac}})|\psi'(t,h_{\rm{ac}}) \rangle - |\langle \psi(t,h_{\rm{ac}})|\psi'(t,h_{\rm{ac}}\rangle)|^2
\end{equation}
where $|\psi'(t,h_{\rm{ac}})\rangle = \partial_{h_{\rm{ac}}} |\psi(t,h_{\rm{ac}}) \rangle$ is the partial derivative with respect to the estimated parameter.

In our analysis, we compare the QFI to $N (2t/\pi)^2$ in order to evaluate the sensor's performance relative to the maximum achievable with non-correlated spins. While $N$ uncorrelated spins can achieve a maximum QFI of $N (2t/\pi)^2$, $N$ correlated spins could overcome this bound till the Heisenberg limit of $N^2 (2 t/\pi)^2$. Thus, the ratio $\text{QFI} / (N (2t/\pi)^2)$ provides insight into how effectively the correlations and coherent lifetimes of the sensor are being utilized.

We first compute the QFI in the ergodic and pFTC phases for varying system sizes. We find that the QFI dynamics is roughly independent of the sensor size in the ergodic phase [Fig.~\ref{QFI_N}(a)].  On the other hand, in the pFTC phase, the QFI increases superlinearly with the system size, with moreover a maximum value reached for   increasingly  long times [Fig.~\ref{QFI_N}(b)]. This time is approximately equal to the lifetime of the pFTC, and the maximum value over time can be interpreted as the best performance of the sensor within a single-shot measurement. We show in Fig.~\eqref{maxQFI} how this best performance depends on the system parameters. Curiously, it does not follow a one-to-one correspondence to the pFTC lifetime, coherence or entanglement features, as observed 
in Fig. ~\eqref{maxQFI}(a).
In the pFTC phase, the maximum QFI value increases with the disorder strength $h$. However, it is not monotonous with the deviation from $\phi=\pi$ kicking phase. We find these results intriguing and warrant further investigation.

We also analyze the impact of nnn and DMI interactions in the sensor - [Fig.~\ref{maxQFI}(b)]. We find that competing $J_2$ are detrimental to the sensor, with the minimum QFI value reached for $\vert J_2/ J_1 \vert \sim 1/2$, indicating again that in this case, frustrated interactions tend to disrupt the (potentially beneficial) order in the system dynamics. Deviations from this ratio, where either $J_1$ or $J_2$ tends to become the dominant interactions, can increase the maximum QFI reached in the single-shot run. Similar to the previous discussion, we also find that DMI interactions have no relevant impact on the sensor. 
Additionally, we calculate the maximum QFI dependence with disorder parameter $h$. In the ergodic phase, the maximum QFI is independent of the system size. However, increasing the disorder deeper in the pFTC phase increases the optimum behavior for a single-shot sensor, with, moreover, a superlinear enhancement with the number of spins [Fig.~\ref{maxQFI}(c)].

\section{Conclusion}
\label{conclusion}
In this work, we investigated the emergence of FTCs in a chiral multiferroic chain and a detailed analysis of its properties; namely, magnetization, entanglement entropy, and coherence. By introducing an AC signal and employing QFI as a tool to quantify the system's sensitivity to the field, we also discussed the system's potential as a quantum sensor.

The analysis of magnetization revealed the two main dynamics regimes of the model, with either an ergodic phase whose magnetization rapidly decays to zero, independent of system size, or a pFTC with a slower decay which could be tuned (or optimized for sensing applications) by the system parameters.  We showed that the lifetime of the pFTC is related to the entropy and coherence saturation values in the system, with lower entropy and coherence saturation values corresponding to longer pFTC lifetimes. 

The QFI calculations provided further insights into the system's performance as a quantum sensor. In the ergodic phase, the QFI is independent of system size, while in the pFTC phase, it shows a superlinear scaling with $N$. The time required to reach the maximum of QFI is aligned to the pFTC lifetime. The system-size dependence highlights how correlations and coherence play a significant role in enhancing the sensor's performance in the pFTC phase, as observed in the ratio to uncorrelated spins, $\text{QFI} / (N t^2)$. Curiously, the maximum of such a ratio over time has not always a direct correspondence to the pFTC lifetime, coherence or entanglement, indicating that additional factors should be taken into account, warranting further investigation. Finally, we analyzed the effects of competing nnn interactions $J_2$ and DMI on the sensor. While frustrating interactions $J_2$ lead to a decrease in the QFI,  DMI stabilizes the formation of chiral spin order. 

These findings provide valuable insights into the understanding of prethermal Floquet time crystals and into the design of quantum sensors based on multiferroic systems, highlighting the role of disorder, correlations, and coherent lifetimes in optimizing its performance.

\section*{ACKNOWLEDGMENTS}

F.I. acknowledges financial support from the Brazilian funding agencies CAPES, CNPQ, and FAPERJ (Grants No. 308205/2019-7, No. E-26/211.318/2019, No. 151064/2022-9, and No. E-26/201.365/2022) and by the Serrapilheira Institute (Grant No. Serra 2211-42166). SKM acknowledges the support provided by the Science and Engineering Research Board (SERB), Department of Science and Technology (DST), India, under the Core Research Grant CRG/2021/007095.

\bibliography{MBL}

\appendix

\section{pFTC lifetime dependence on system parameters}
\label{appendix}

\begin{figure}
\includegraphics[width=.95\linewidth, height=.65\linewidth]
{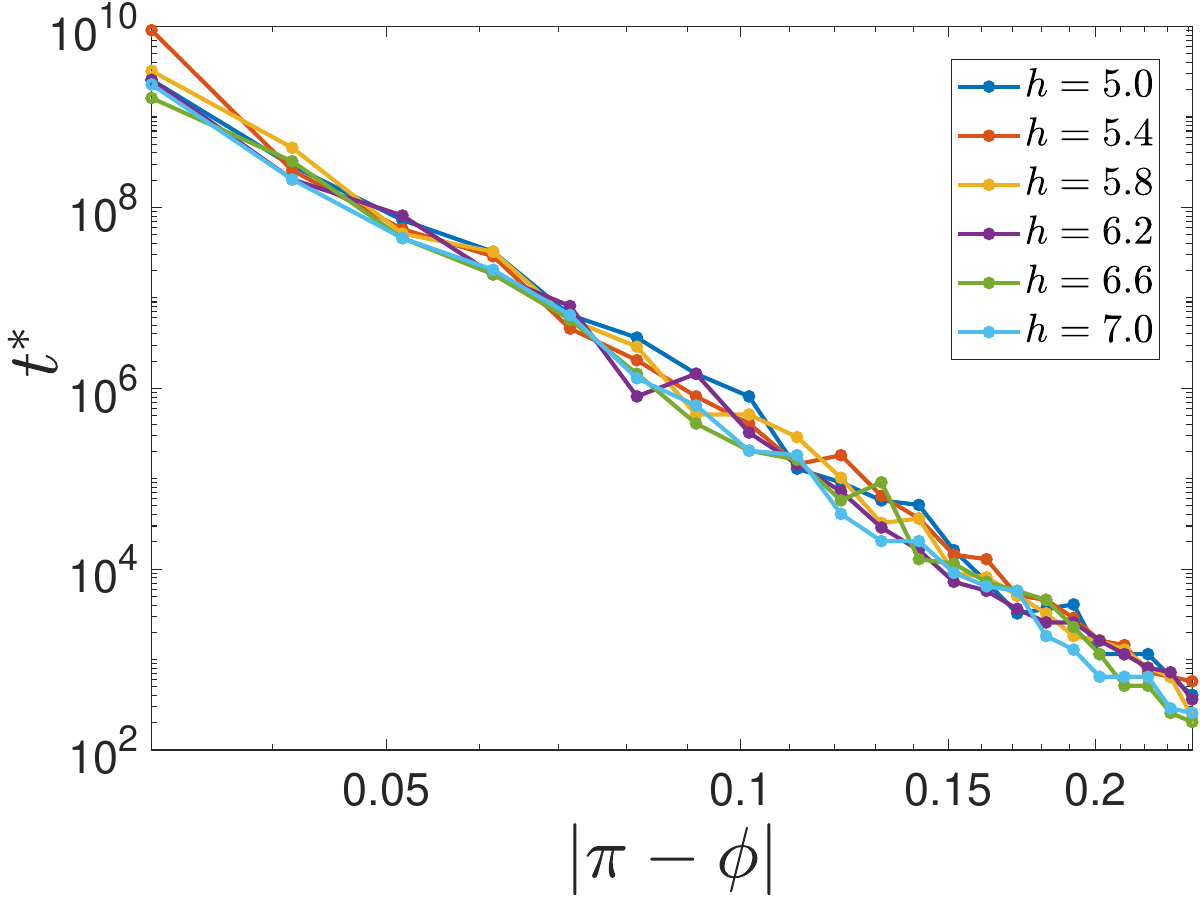}
\caption{Lifetime of the pFTC as a function of $\vert \pi-\phi \vert$ for different values of $h$ in a log-log scale. The parameters used are $J_1 = -1$, $J_2 = -J_1/4$, $D = 0$, and $N = 8$, with a total of 250 disorder realizations.}
   \label{t_phi}
  \end{figure}  

In this appendix we study the dependence of the pFTC lifetime on the system parameters, specifically, its dependence with the kicking phase $\phi$ and strength of disordered magnetic field, $h$.
We obtain that the lifetime increases as one approaches the perfect $\pi$-kick phase as a power-law,  
$t^{*} \sim \vert \pi-\phi \vert^{-\gamma} $
where $\gamma > 0$ - 
see Fig.\eqref{t_phi}. It is important to recall that the model at $\phi=\pi$ is a 
fine-tuning point where since the total magnetization ($\hat S_z(t)$) is a conserved quantity in the bare $\hat H_0$ Hamiltonian, for perfect
kick flips one has a constant magnetization at even stroboscopic time steps. In other words, in this case the period doubling is not representative of the phases, which would have an ``infinite lifetime" regardless of the underlying pFTC or ergodic microscopic spin dynamics within the conserved subspaces.

\begin{figure}
\includegraphics[width=.95\linewidth, height=.65\linewidth]
{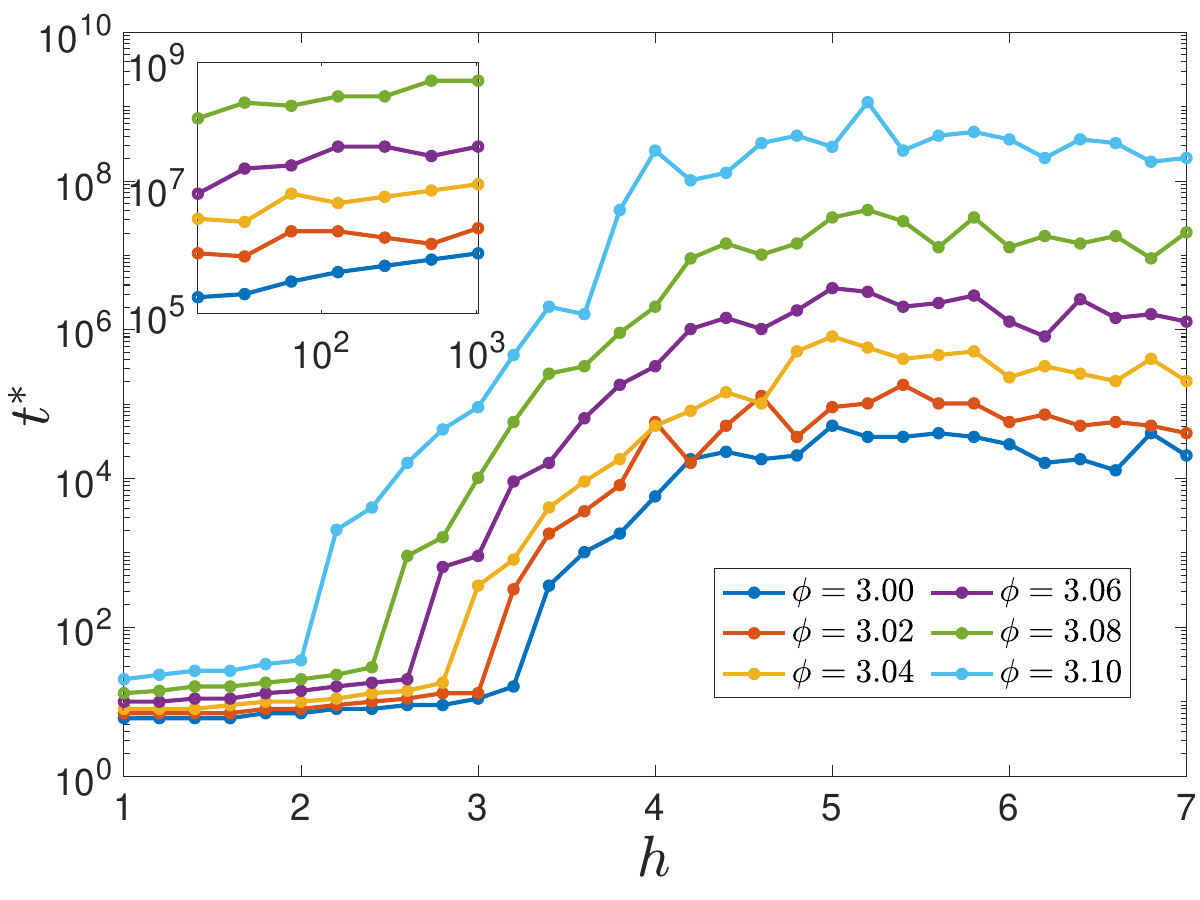}
\caption{Lifetime of the pFTC as a function of $h$ for different values of $\phi$ in a log-linear scale. The parameters used are $J_1 = -1$, $J_2 = -J_1/4$, $D = 0$, and $N = 8$, with a total of 250 disorder realizations.} 
   \label{t_h}
  \end{figure}  
The lifetime dependence with the disorder strength shows different scalings - see Fig.\eqref{t_h}. While in the ergodic phase it is roughly independent on the disorder, as one enters the pFTC regime we observe a rapid increase in the lifetime with \( h \), which approximately follows an exponential growth \( t^{*} \sim e^{\beta h} \) with $\beta>0$. For more stable kicking phases \( \phi \to \pi \), the exponential growth becomes even more pronounced, providing stronger evidence for this behavior. For increasingly larger disorder stengths we observe a further different scaling, where the lifetime now increases as a power-law with the disorder strength - see inset of Fig.(\ref{t_h}).

\end{document}